\documentclass[a4paper,11pt]{amsart}
\begin{document}
\hyphenation{gra-vi-ta-tio-nal re-la-ti-vi-ty Gaus-sian
re-fe-ren-ce re-la-ti-ve gra-vi-ta-tion Schwarz-schild
ac-cor-dingly gra-vi-ta-tio-nal-ly re-la-ti-vi-stic pro-du-cing
de-ri-va-ti-ve ge-ne-ral}
\title[No motions of bodies produce GW's]
{{\bf  No motions of bodies produce GW's}}
\author[Angelo Loinger]{Angelo Loinger}
%\date{}
\address{Dipartimento di Fisica, Universit\`a di Milano, Via
Celoria, 16 - 20133 Milano (Italy)}
\email{angelo.loinger@mi.infn.it}
\thanks{To be published on \emph{Spacetime \& Substance.}}

\begin{abstract}
A close comparison between Maxwell field and Einstein field makes
conceptually and immediately evident that in general relativity
(GR) no motions of bodies can generate gravitational waves (GW's).
\end{abstract}

\maketitle

%%\begin{equation} \label{eq:sevenprime}
%%    \ddot{\Re} + \frac{\kappa}{6}\Re \rho=0 , \tag{7'}
%% \end{equation}
%% ``mechanisms'' \textrm{d} \`a
%% \cite{1}

\vskip1.20cm \noindent \textbf{1}. -- Let us consider a continuous
``cloud of dust'' characterized by a material energy tensor $\rho
u^{j}u^{k}$, $(c=1)$, $(j,k=0,1,2,3)$; $\rho$ is the invariant
mass density and $u^{j}$ is the four-velocity. We assume first
that this ``dust'' is electrically charged -- with an invariant
charge density $\sigma$ --, and that the gravitational interaction
between its particles is negligible. Thus we have a total mass
tensor $T^{jk}$ given by

\begin{equation} \label{eq:one}
T^{jk}=\rho u^{j}u^{k}+S^{jk}\quad,
\end{equation}

where $S^{jk}$ is the energy tensor of Maxwell field. Suppose that
the spatio-temporal substrate is a Minkowskian manifold, which is
referred to a system of \emph{general} co-ordinates
$x^{0},x^{1},x^{2},x^{3}$. If a colon denotes a covariant
differentiation, we have, as it can be formally proved (see sect.
\emph{\textbf{A.2}} of \emph{\textbf{Appendix A}}):

\begin{equation} \label{eq:two}
%  T^{jk}_{\,:k}=0 \quad,
%  T^{jk}_{\::k}=0 \quad,
%  T^{jk}_{\;:k}=0 \quad,
%  T^{jk}_{\ :k}=0 \quad,
  T^{jk}_{\quad:k}=0 \quad,
%  T^{jk}_{\qquad:k}=0 \quad,
\end{equation}

i.e. the differential conservation law of tensor $T^{jk}$.

\par From

\begin{equation} \label{eq:three}
4\pi S^{jk}:= -F^{j}_{\ r}F^{kr}+ \frac{1}{4}f^{jk}F_{rs}F^{rs}
\quad,
\end{equation}

where $F^{jk}$ is the e.m. field and $f^{jk}$ the metric tensor
--, taking into account Maxwell equations

\begin{equation} \label{eq:four}
F_{jk:r} +  F_{kr:j} +F_{rj:k} = F_{jk,r} +  F_{kr,j} +F_{rj,k} =0
\quad,
\end{equation}

(the comma denotes ordinary differentiation), and

\begin{equation} \label{eq:five}
F^{jk}_{\quad:k} = 4\pi \sigma u^{i} \quad,
\end{equation}

\newpage
we obtain

\begin{equation} \label{eq:six}
S^{jk}_{\quad:k} = 4\pi F^{jk}\sigma u_{k} \quad.
\end{equation}

Thus eqs.(\ref{eq:two}) give

\begin{equation} \label{eq:seven}
\rho u^{j}u^{k}_{\ :j} + F^{kj}\sigma u_{j}=0  \quad,
\end{equation}

which represent the equations of motion of the charged particles;
they are an \emph{analytical consequence} of differential
conservation eqs.(\ref{eq:two}) and of Maxwell equations
(\ref{eq:four}) and (\ref{eq:five}). \cite{1}.

\par The theoretical existence of the electromagnetic waves is an
analytical consequence of the above equations, as it is known.

\vskip0.50cm \noindent \textbf{2}. -- We consider now a ``dust'',
the particles of which interact \textbf{\emph{only}}
gravitationally; a physical example: the solar system. According
to GR we have ($G=1$):

\begin{equation} \label{eq:eight}
R^{jk}-\frac{1}{2}g^{jk}R = -8\pi \rho u^{j}u^{k}  \quad;
\end{equation}

by virtue of Bianchi relations, the (left-hand side)$_{:k}$
vanishes identically; consequently:

\begin{equation} \label{eq:nine}
(\rho u^{j}u^{k})_{:k}=0 \quad,
\end{equation}

from which:

\begin{equation} \label{eq:ten}
(\rho u^{j})_{:j}=0 \quad,
\end{equation}

i.e. the mass conservation, and

\begin{equation} \label{eq:eleven}
u^{j}u^{k}_{\ :j}=0 \quad,
\end{equation}

i.e. the \textbf{\emph{geodesic}} equations of motion. Clearly,
the geodesic motions \textbf{\emph{cannot}} generate GW's.

\par Remark the fundamental difference with Maxwell case of
sect.\textbf{1.}: in lieu of eqs.(\ref{eq:six}), we have here:

\begin{equation} \label{eq:twelve}
(R^{jk}-\frac{1}{2}g^{jk}R)_{:k} = 0  \quad,
\end{equation}

i.e. relations which do not involve explicitly the four-velocity
$u^{j}$. Accordingly, in lieu of eqs.(\ref{eq:seven}), we have the
simple geodesic eqs.(\ref{eq:eleven}).

\vskip0.50cm
\noindent \textbf{3}. -- For an electrically charged
``dust'', the particles of which interact gravitationally (and
electromagnetically), we obtain the following equations of motion:

\begin{equation} \label{eq:thirteen}
\rho u^{j}u^{k}_{\ :j} + F^{kj}\sigma u_{j}=0 \quad,
\end{equation}

i.e. four equations which are \emph{formally} analogous to
eqs.(\ref{eq:seven}). Here the motions are \emph{not} geodesic.
However, \emph{no} GW can be emitted, for the following reason.

\par Assume, for simplicity's sake only, that $\rho$ vanishes
everywhere in spacetime, except for a thin tube of world lines.
Suppose further that at a given time $t=t'$ the particle $P$
extending over the tube begins to emit GW's; let $K'(t')$ be the
set of kinematical elements (velocity, acceleration, time
derivative of the acceleration, \emph{etc}.) of $P$ at $t=t'$.
Consider now another ``dust'', identical to the previous one, but
such that its charge density $\sigma$ is equal to \emph{zero}, and
immerse it in a \emph{suitable} ``external'', ``fixed''
gravitational field. It is obvious that at some time $t''$
particle $P$ will have a set of kinematical elements, say
$K''(t'')$, which is equal to the above $K'(t')$. But now the
particle $P$ describes a tube of \emph{geodesic} lines, and
therefore \emph{no} GW can be emitted.

\par We see in particular that the current conviction according to
which an accelerated mass sends forth GW's is false. This belief
was originated in the old times by a \textbf{\emph{partial}}
analogy between Maxwell theory and the \textbf{\emph{linearized}}
version of GR. \cite{1}.

\vskip0.50cm \noindent \textbf{4}. -- The line of reasoning of
previous sect.\textbf{2.} can be extended to the case of a
``dust'' characterized by a mass tensor $T^{jk}=(\mu +
p)u^{j}u^{k}+pg^{jk}$, where $\mu$ and $p$ are scalars connected
by an equation of state $\mu = \varphi (p)$ \cite{2}. The
differential conservation relations of GR

\begin{equation} \label{eq:fourteen}
T^{jk}_{\quad:k}=0
\end{equation}

give the Eulerian equations of motion of perfect fluid
hydrodynamics. Obviously, the motions of the particles are
\emph{not} geodesic; however, with an argument similar to that of
sect.\textbf{2.} we may conclude that \emph{no} emission of GW's
is possible.

\vskip0.50cm \noindent \textbf{5}. -- In Maxwell theory the
divergence of the e.m. energy tensor is zero only in the absence
of charges and currents; in general, it is equal to $4\pi
F^{jk}J_{k}$, cf. eqs.(\ref{eq:six}). On the contrary, in GR we
have the fundamental circumstance that the divergence of the
left-hand side of (\ref{eq:eight}) is always (identically) equal
to zero:

\begin{equation} \label{eq:fifteen}
(R^{jk}-\frac{1}{2}g^{jk}R)_{:k} = 0  \quad;
\end{equation}

in the last analysis, the absence of a ``mechanism'' for the
emission of GW's can be ascribed to eqs.(\ref{eq:fifteen}). --

\par Consider, quite generally, \emph{a generic continuous
medium}; the differential conservation equations

\begin{equation} \label{eq:sixteen}
T^{jk}_{\quad:k}=0
\end{equation}

yield -- completely or partially -- its equations of motion, which
in general are not geodesic. However, the kinematical eòlements of
these motions are not different from the kinematical elements of
purely geodesic motions, and therefore \emph{no} GW is sent forth
-- not even by catastrophic astrophysical perturbations.

\par The mass tensor $T^{jk}$ is the sum of a material (\emph{stricto
sensu}) energy tensor and of the energy tensors of all the fields,
\emph{except} the metric tensor $g_{jk}$, which is the substance
of spacetime. This very special character of $g_{jk}$ is
responsible for the fact that the undulatory solutions of Einstein
equations are destitute of a physical reality.

\vskip0.50cm \noindent \textbf{5}. -- We have seen that, contrary
to what occurs in Maxwell theory for the e.m. waves, in GR no
motions of bodies can give origin to GW's. In my opinion, the
previous considerations are rather stringent. Moreover, their
correctness is indirectly confirmed by various other proofs of the
same result \cite{3}. There are many roads to Rome.

\par Unfortunately, the astrophysical community is still very far
from \emph{Caput mundi}, and has entered into a blind alley
\cite{4}. Getting rid of current mythological ideas on GW's -- and
BH's -- will be a painful operation.

\vskip0.70cm
\begin{center}
\noindent \small \emph{\textbf{APPENDIX A}}\normalsize
\end{center}

\vskip0.20cm \noindent \emph{\textbf{A.1.}} -- Let us consider the
two following integrals, $I$ and $I'$, over a generic
spatio-temporal region $D$:

\begin{equation} \label{eq:seventeen}
I=\int_{D} R\sqrt{-g} \textrm{ d}x \quad,
\end{equation}

with $R=R^{jk}g_{jk}$, and $\textrm{d}x \equiv
\textrm{d}x^{0}\textrm{d}x^{1}\textrm{d}x^{2}\textrm{d}x^{3}$, and

\begin{equation} \label{eq:eighteen}
I'=\int_{D} \mathcal{L}\textrm{ d}x \quad,
\end{equation}

where $\mathcal{L}$ is a scalar density, which does not contain
derivatives of the $g_{jk}$ -- for simplicity's sake.
$\mathcal{L}$ is a function of the physical quantities of the
system -- as four-velocities, e.m. fields, hydrodynamical
observables, \emph{etc}. -- and of their derivatives.
$\mathcal{L}$ must be such that

\begin{equation} \label{eq:nineteen}
\frac{\partial \mathcal{L}}{\partial g_{jk}} = T^{jk} \sqrt{-g}
\quad.
\end{equation}

Consider now the variations of $I$ and $I'$, say
$\delta_{\textrm{\large*\normalsize}}I$ and
$\delta_{\textrm{\large*\normalsize}}I'$, generated by a variation
$\delta_{\textrm{\large*\normalsize}}g_{jk}$ of the metric tensor,
which is induced by an \emph{infinitesimal} co-ordinate
transformation

\begin{equation} \label{eq:twenty}
x'^{j} = x^{j}+ \epsilon^{j}(x) \quad,
\end{equation}

such that the functions $\epsilon^{j}(x)$ vanish at the bounding
surface of $D$.

\par By well known computations \cite{5}, the invariance
conditions
$\delta_{\textrm{\large*\normalsize}}I=\delta_{\textrm{\large*\normalsize}}I'=0$
yield respectively:

\begin{equation} \label{eq:twentyone}
(R^{jk}-\frac{1}{2}g^{jk}R)_{:k} = 0  \quad,
\end{equation}

\begin{equation} \label{eq:twentytwo}
T^{jk}_{\quad:k}=0 \quad,
\end{equation}

i.e. the differential conservation equations of the tensor
$[R^{jk}-(1/2)g^{jk}R]$ and of the mass tensor $T^{jk}$.

\par Remarkable facts: \emph{i}) the physical results
(\ref{eq:twentyone}) and (\ref{eq:twentytwo}) are a mere
consequence of the above \emph{formal} invariance; \emph{ii}) they
have been deduced in a way that is \emph{fully independent} of
Einstein equations

\begin{equation} \label{eq:twentythree}
R^{jk}-\frac{1}{2}g^{jk}R = -8\pi T^{jk}  \quad;
\end{equation}

\emph{iii}) eqs.(\ref{eq:twentyone}) have been deduced
\emph{without} using Bianchi relations.

\vskip0.50cm \noindent \emph{\textbf{A.2.}} -- Another remarkable
fact is the following. Consider the case for which the
gravitational interactions are negligible, and consider a
formulation of SR in \emph{arbitrary} co-ordinates
$x^{0},x^{1},x^{2},x^{3}$.

\par The differential conservation equations for the mass tensor $T^{jk}$ can
be derived in this way: choose, as in \textbf{\emph{A.1.}}, a
scalar density $\mathcal{L}$ such that

\begin{equation} \label{eq:tewntyfour}
\frac{\partial \mathcal{L}}{\partial g_{jk}} = T^{jk} \sqrt{-f}
\quad,
\end{equation}

where $f:=\det \|f_{jk}\|$, and $f_{jk}(x)$ is the metric tensor;
the condition $\delta_{\textrm{\large*\normalsize}}I'=0$, induced
by $\delta_{\textrm{\large*\normalsize}}f_{jk}$, gives -- exactly
as in \textbf{\emph{A.1.}} --:

\begin{equation} \label{eq:twentyfive}
T^{jk}_{\quad:k}=0 \quad.
\end{equation}

\par This means that, contrary to a diffuse opinion, \textbf{\emph{also in
SR}} eqs.(\ref{eq:twentyfive}) are a mere consequence of a simple
invariance property of $I'$. The merit of this outcome pertains
\textbf{\emph{only}} to the formulation in \textbf{\emph{general}}
co-ordinates $x^{0},x^{1},x^{2},x^{3}$.

\vskip0.50cm \noindent \emph{\textbf{A.3.}} -- The equations of
motion of a considered material continuum can be derived
(completely or partially) from eqs.(\ref{eq:twentytwo}) in GR and
from eqs.(\ref{eq:twentyfive}) in SR.

\par \emph{The nonlinearity of Einstein
eqs.(\ref{eq:twentythree}) has nothing to do with this fundamental
result} -- contrary to a widespread belief.

\vskip0.50cm \noindent \emph{\textbf{A.4.}} --  A final remark. I
have emphasized that eqs.(\ref{eq:twentyone}) have been here
derived without using Bianchi identities. However, the present
deduction \emph{depends} on the relativistic \emph{dichotomy}
between the gravitational potential $g_{jk}$ and the other fields.
Otherwise, in lieu of the \emph{two} conditions
$\delta_{\textrm{\large*\normalsize}}I=0$ and
$\delta_{\textrm{\large*\normalsize}}I'=0$, we ought to write the
\emph{unique} (and weaker) condition
$\delta_{\textrm{\large*\normalsize}}(I+I')=0$.

\newpage
%\vskip0.70cm
\begin{center}
\noindent \small \emph{\textbf{APPENDIX B}}\normalsize
\end{center}

\vskip0.20cm Let us reconsider the case (see sect.\textbf{1}.) of
a continuous charged ``dust'' -- whose particles (of finite size)
interact \textbf{\emph{only}} electromagnetically --, when the
Minkowskian spacetime is described by the customary metric tensor
$\eta_{jk}$ such that: $\eta_{rs}=0$ for $r\neq s$, $\eta_{00}=1$,
 $\eta_{11}=\eta_{22}=\eta_{33}=-1$.

\par A pedestrian repetition of the reasoning of sect.\textbf{1}.
-- with the only substitution of the ordinary differentiation for
the covariant one -- leads us to conclude that the equations of
motion of the charged particles

\begin{equation} \label{eq:twentysix}
\rho u^{j}u^{k}_{\ ,j} + F^{kj}\sigma u_{j}=0
\end{equation}

are a mere consequence of the \emph{definition} of the mass tensor
$T^{jk}$ $[$eqs.(\ref{eq:one})$]$ and of Maxwell field equations.
This result is generally ignored in the treatises dealing the
electromagnetic theory. On the contrary, in the standard
formulations of Maxwell electrodynamics it is affirmed that the
law of motion of the charges is independent of the field
equations.

\par It is interesting that \emph{both in GR and in SR} the
equations of motion of \emph{any} material continuum are
\emph{only} analytical consequences of the \emph{definition} of
the mass tensor $T^{jk}$ and of Einstein and/or Maxwell
\emph{field} equations.

\vskip0.70cm
\begin{center}
\noindent \emph{\textbf{\emph{Parergon}}}
\end{center}

\vskip0.20cm The searchers of GW's and of BH's have entered into a
\emph{cul-de-sac}, owing to their reluctance to abandon erroneous
\emph{loci communes} concerning the real physical meaning of GR. A
conceptually inadequate ``Vulgate'', based on second-hand works,
has got the upper hand. The papers quoted in \cite{4} are an
example of this situation. (\emph{Stat pro ratione libido}).

\par From the ``Conclusions'' of the first paper in \cite{4}:
``Two different astrophysical searches were performed: an all-sky
search aimed at signals from isolated neutron stars and an orbital
parameter search aimed at signals from the neutron star in the
binary system ScoX-1. Both searches also cover a wide range of
possible emission frequencies: a 568.8 Hz band for the isolated
pulsar search and two 20-Hz bands for the ScoX-1 search. -- The
sensitivity of these analyses makes the detection of a signal
extremely unlikely. As a consequence the main goal of the paper is
to demonstrate an analysis method using real data $[$\ldots$]$''
(An implicit admission of another fiasco).

\par From p.54 of the second paper in \cite{4}:
``The coalescence of two relativistic stars (double neutron star
or black hole/neutron star binary mergers) is the end result of
0.1-1 Gyr of orbital decay caused by the emission of gravitational
waves. This paroxysmal event should also give rise to a black hole
surrounded by a torus of matter at nuclear densities, possibly
producing relativistic jets that are less energetic and shorter
lived than those of collapsars and originating short GRBs.'' (A
report from the dream-land: a bundle of unfounded conjectures).


\vskip0.70cm \small
\begin{thebibliography}{9}

\bibitem{1}
See also: H. Weyl, \emph{Amer. J. Math.}, \textbf{66} (1944) 591;
A. Loinger, \emph{Spacetime \& Substance}, Vol.\textbf{5}, No.2
(22), 2004, p.53.

\bibitem{2}
See e.g. V. Fock, \emph{The Theory of Space, Time and Gravitation}
(Pergamon Press, Oxford, \emph{etc}.) 1964, sect.\textbf{32} and
Appendix \textbf{\emph{C}}.

\bibitem{3}
Cf. A. Loinger, \emph{arXiv:physics/0603214 v1} (March 25th, 2006)
-- in course of publication on \emph{Spacetime \& Substance} --
\textbf{\emph{and references therein}}.

\bibitem{4}
See e.g.: The LIGO Scientific Collaboration,
\emph{arXiv:gr-qc/0605028 v1} (May 4th, 2006); G. Chincarini
\emph{et al.}, \emph{The Messenger}, \textbf{123} (March 2006) 54.

\bibitem{5}
Cf. e.g.: E. Schr\"odinger, \emph{Space-Time Structure} (Cambridge
University Press, Cambridge) 1960, p.93 \emph{seqq.}; P.A.M.
Dirac, \emph{General Theory of Relativity} (J. Wiley and Sons, New
York, \emph{etc}.) 1975, pp.59 and 60. See also: A. Loinger,
\emph{Nuovo Cimento}, \textbf{A110} (1997) 341; Idem,
\emph{ibidem}, \textbf{A112} (1999) 407.

\end{thebibliography}
\end{document}